# Redox and Molecular Diffusion in 2D van der Waals Space

Haneul Kang,[#] Kwanghee Park[#] and Sunmin Ryu*

Department of Chemistry, Pohang University of Science and Technology (POSTECH), Pohang, Gyeongbuk 37673, Korea

* E-mail: sunryu@postech.ac.kr

## Abstract

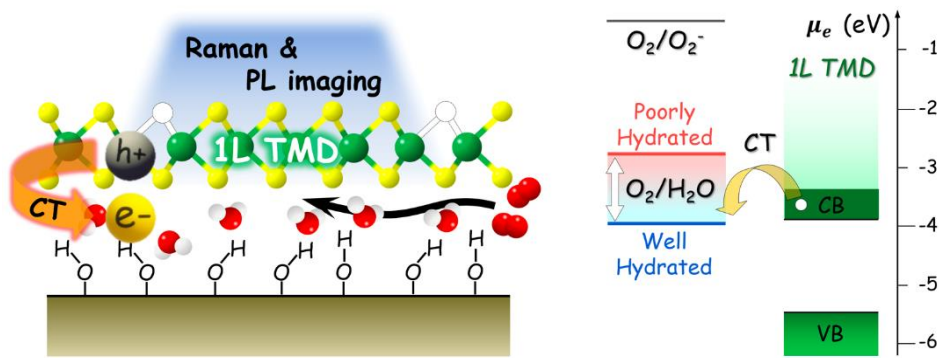

Understanding charge transfer (CT) between two chemical entities and subsequent change in their charge densities is essential not only for molecular species but also for various low-dimensional materials. Because of their extremely high fraction of surface atoms, two-dimensional (2-D) materials are most susceptible to charge exchange and exhibit drastically different physicochemical properties depending on their charge density. In this regard, spontaneous and uncontrollable ionization of graphene in the ambient air has caused much confusion and technical difficulty in achieving experimental reproducibility since its first report in 2004. Moreover, the same ambient hole doping was soon observed in 2-D semiconductors, which implied that a common mechanism should be operative and apply to other low-dimensional materials universally. In this Account, we review our breakthroughs in unraveling the chemical origin and mechanistic requirements of the hidden CT reactions using 2-D crystals. We developed in-situ optical methods to quantify charge density using

Raman and photoluminescence (PL) spectroscopy and imaging. Using gas and temperature-controlled in-situ measurements, we revealed that the electrical holes are injected by the oxygen reduction reaction (ORR): $O_2 + 4H^+ + 4e^- \rightleftarrows 2H_2O$, which was independently verified by pH dependence in HCl solutions. In addition to oxygen and water vapor, the overall CT reaction requires hydrophilic dielectric substrates, which assist hydration of the sample-substrate interface. The interface-localized reaction allowed us to visualized and control interfacial molecular diffusion and CT by varing the 2-D gap spacing and introducing defects. The complete mechanism of the fundamental charge exchange summarized in this Account will be essential in exploring material and device properties of other low dimensional materials.

# 1. Introduction

**Charge transfer doping by oxygen-water redox couple.** Charge transfer (CT) between two connected chemical entities causing redox reactions is a ubiquitous phenomenon occurring in varying length scales that range from a single pair of molecules forming a CT complex[1] to two metal electrodes leading to built-in contact potential difference.[2] CT also underlies chemical doping (or surface CT doping) of conductive polymers,[3] which proved to be effective and non-destructive unlike substitutional doping used in conventional semiconductor devices. Because of its interfacial nature, chemical doping using halogens and alkali metals found wide application in various low-dimensional materials with a high surface-atom percentage such as carbon nanotubes (CNTs),[4-5] graphene,[6-8] and transition metal dichalcogenides (TMDs).[9] Chemical modulation in charge density led to a drastic change in their electronic structures, consequently affecting their electrical and optical properties.

Unlike such simple two-body exchanges by single-entity electron donors or acceptors, however, CT involving dioxygen and water has long been a controversy and caused significant confusion over the origin of ambient hole doping in various low-dimensional materials (Fig. 1a & 1b). Spontaneous[10-11] and thermally activated[12-14] ionization of graphene discovered in the ambient air by electrical and optical means (Fig. 1c) remained elusive and controversial regarding their mechanisms. The same air-mediated hole doping was observed in two-dimensional (2-D) semiconductors such as $MoS_2$ and $WS_2$ but without a clear mechanistic understanding (Fig. 1d).[15-16] A similar debate was made over the mechanism of the air-enhanced conductivity of CNTs.[17-19] As CNTs and their junctions with metal electrodes both were sensitive to $O_2$, electrical measurements could not readily provide a decisive conclusion between $O_2$-mediated hole doping and reduction of the Schottky barrier.[20] In addition, the role of water in the enhanced conduction was much less explored than $O_2$, requiring systematic and definitive scrutiny.[17] An almost parallel controversy occurred even earlier among those who observed air-enhanced surface conductivity for hydrogen-terminated diamonds since 1989.[21] Despite several mechanisms proposed for the apparent hole injection in diamonds, a complete explanation[22] had to wait for nearly two decades.

**Relation to ORR and refinements required.** A set of pioneering studies implied that the accepting half of the elusive redox process is the oxygen reduction reaction (ORR): $O_2 + 4H^+ + 4e^- \rightleftarrows 2H_2O$,[13, 22-26] where electrons are taken from the above low-dimensional materials. The recurring controversy mentioned above, however, implies that there are complex issues in applying the mechanism. In 2007, J. Angus and coworkers experimentally validated that ORR is responsible for the unusually high surface conduction of hydrogenated diamonds,[22] the electrochemical nature of which was first proposed several years earlier.[27] In the Marcus-Gerischer (MG) model,[28] the redox can be described as CT between $O_2/H_2O$ redox couples and hydrated solid electron donors (Fig. 1e). The CT rate depends on the energetic alignment between the donor's Fermi level and acceptor's oxidized state,[23] the latter of which should depend on pH and oxygen concentration.

However, we note a few crucial factors that need to be considered for complete understanding. For a solid donor exposed to a gaseous environment, the aqueous phase required for the electrochemical reaction is not well defined and spatially limited to the donor's surface. In particular, the hydration level of the donor and acceptor varies substantially and is generally insufficient to invoke conventional electrochemical redox potentials. In this regard, the physical contact with hydrophilic substrates (Fig. 1b) can play a pivotal role in modulating CT for low-dimensional materials, as seen in CNT.[29-30] Another difficulty is its slow and sometimes widely varying kinetics originating from its reversibility,[23] which complicates experimental characterization. In addition, as this particular ORR proceeds in an open circuit, the Fermi level of the electrically floated donor is not constant but varies in the direction favoring the reversed reaction, eventually reaching an equilibrium.[26] Lastly, the nature of CT reaction centers is far from being understood, although they are crucial in the ORR kinetics.

**Optical visualization of redox and molecular diffusion in 2-D space.** Through a series of works,[12-13, 24-26] we established that the above ORR in Fig. 1e is responsible for the air-induced hole doping in 2-D metals and semiconductors. The optical investigation of dielectric-supported 2-D crystals has a few unique strengths in studying their CT processes compared with other forms of materials. Because of their high surface-atom percentage, even a tiny change in areal charge density (~$10^{11}$ cm$^{-2}$) induces significant perturbation on the overall vibrational and excitonic behaviors, which can be detected with Raman (Fig. 1c) and photoluminescence (PL; Fig. 1d) spectroscopies, respectively. Their high crystallinity and physicochemical tunability of defects allow reproducible and systematic investigation. The nanometer-scale gap between the impermeable 2-D crystals and underlying dielectric substrates serves as an ideal 2-D space that accommodates and confines the molecular redox species. The chemical and geometric details of the substrates can also be tuned as variables in modulating CT. Lastly, the 2-D space presents an opportunity to study molecular diffusion in constricted space as molecular species are transported through the nanoscopic space during the ORR-driven CT process.

In this Account, we review our experimental breakthroughs in optical imaging of redox and molecular diffusion in 2-D van der Waals (vdW) space. By establishing quantitative spectroscopic and imaging probes for charge density, we verified that the chemical origin of the spontaneous hole doping in 2-D crystals is ORR, which is localized at the sample-substrate interface. Our study showed that the nanoscopic electrochemical reaction is mediated by interfacial water and can be modulated by varying the hydrophilicity of substrates. We also demonstrated direct visualization and manipulation of molecular diffusion and subsequent CT by modifying the 2-D space geometrically and chemically. Our findings providing a complete mechanism for the fundamental charge exchange will be essential in exploring material and device properties of other low dimensional materials.

## 2. Development of optical probes for charge density

The most apparent challenge and origin of the confusion were that the CT reaction involves $O_2$ and water ubiquitously present in the ambient conditions. Unless specified otherwise hereafter, CT will refer to electron transfer from 2-D materials to the molecular species. Moreover, the amount of consumed reactants is a few orders of magnitude smaller than a monolayer coverage. The typical charge density (denoted $n$ in general, $n_e$ and $n_h$ specifically for electron and hole, respectively) induced by various chemical dopants is on the order of $10^{13}$ cm$^{-2}$, which translates into one electrical hole or a quarter of $O_2$ molecule per ~400 carbon atoms in graphene or ~120 unit cells in 1L $MoS_2$. Despite the small fraction, however, the extra charges readily affect electrical conductivity,[11] electronic transitions[31] and lattice vibrations[32] of various 2-D materials. Whereas various forms of charge density probes could be devised based on this fact, we selected to develop sensitive optical methods. As explained below, Raman and PL spectroscopies allowed us to resolve a fraction of $10^{12}$ cm$^{-2}$ in graphene and TMDs, respectively. Unlike electrical methods, optical probes do not require any pretreatments of samples and also provide high spatial resolution only limited by diffraction.

Like other carbon materials, graphene exhibits significant changes in its lattice vibration upon external perturbation such as charge injection,[32-33] lattice strain[34] and defects.[35] Because of this fact combined with the inherent strengths of inelastic scattering, Raman spectroscopy became a popular characterization tool for graphene[36] and other 2-D materials, including TMDs.[37] However, the multimodal sensitivity of Raman peak frequencies, $\omega_G$ (G mode) and $\omega_{2D}$ (2D mode) for graphene, hampered quantifying the degree of each perturbation.[25] In particular, their unique geometry with an extreme aspect ratio between length and thickness renders 2-D material samples prone to structural distortion when supported on solid substrates.[38] To resolve the complication, J. Lee et al.[25] devised a Raman peak analysis scheme that enables simultaneous determination of $n_h$ and lattice strain ($\varepsilon$) from a single Raman spectrum of graphene. They exploited the fact that $\omega_G$ and $\omega_{2D}$ exhibit differing dependence on $n$[39] and $\varepsilon$,[40] as can be seen in the $\omega_G$-$\omega_{2D}$ frequency correlation (Fig. 2a). Defining that the origin of $\omega_G$-$\omega_{2D}$ space, $O(\omega_G^0, \omega_{2D}^0)$, represents intrinsic graphene, the frequency coordinate of a given sample, $P(\omega_G, \omega_{2D})$, will be displaced along unique trajectories depending on the nature of perturbation. Lattice distortion due to uniaxial stress and hole doping induce displacement along $\varepsilon$ (blue line) and $n_h$ (red line) axes with a slope of $(\Delta\omega_{2D}/\Delta\omega_G)_\varepsilon^{uniaxial} = 2.2 \pm 0.2$ and $(\Delta\omega_{2D}/\Delta\omega_G)_{n}^{hole} = 0.70 \pm 0.05$, respectively. Assuming that the two perturbations are independent of each other with a few approximations, $P(\omega_G, \omega_{2D})$ can be vectorially decomposed into $\varepsilon$ and $n_h$ as depicted in Fig. 2a.[25] With the spectral precision of 1 cm$^{-1}$, one can attain high resolution of ~0.02% in $\varepsilon$ and ~1x10$^{12}$ cm$^{-2}$ in $n_h$. Whereas the original analytic scheme was based on experimental data in the range of -0.6 ~ 0.3% ($\varepsilon$) and 0 ~ 1.6x10$^{13}$ cm$^{-2}$ ($n$), further calibration will allow quantification in wider ranges.

Because of the dependence of $\omega_{2D}$ on the electronic band structure of graphene, the above analysis could be extended to additional functionality. When graphene is weakly coupled with other materials through vdW bonding, the slope (Fermi velocity $v_F$) of its linear $\pi$ bands is decreased because of the electronic coupling.[41] In general, the doubly resonant scattering of 2D mode[42] leads to the upshift of $\omega_{2D}$, not $\omega_G$, for reduced $v_F$,[43] which allows one to determine the degree of electronic coupling. When graphene contacts hexagonal BN[44] and Cu film,[45] $\omega_{2D}$ increases by 6.5 ~ 17 cm$^{-1}$, which corresponds to $\Delta v_F/v_F$ (black line) of 2.9 ~ 14.6% as shown in Fig. 2a. Therefore, neglecting this effect leads to a significant error in the quantification of $\varepsilon$ and $n_h$. We also extended the metrology to other popular excitation wavelengths, which is required because the double-resonant scattering induces significant dispersion in $\omega_{2D}$, unlike $\omega_G$.[46] The charge density, strain, and $O(\omega_G^0, \omega_{2D}^0)$ were calibrated for 457, 514 and 633 nm using chemical doping with $H_2SO_4$, native strain and suspended graphene, respectively (Fig. 2b).[47] Additionally, the quantification of $\Delta v_F/v_F$ in as-grown graphene on Cu was demonstrated for the three wavelengths.[45] We also note that the strain analysis could also be made for multilayer graphene, as shown in Fig. 2c.[48]

For TMDs, we found that PL spectroscopy is more sensitive and versatile than Raman scattering. In principle, A and E phonon modes allow straightforward determination of the charge density[49] and strain[50] in TMDs, respectively, because the two quantities affect the two modes almost exclusively. However, the frequency shift per unit charge density was several times smaller than that of graphene,[25] leading to a less satisfactory resolving power for $n$. Spatial mapping of charge density would also require a long measurement time to collect Raman spectra in a raster scanning manner as done for graphene.[24] In comparison, the ratiometric intensity analysis of PL signals from excitons and charged excitons (trions)[51] could be used to determine $n$ with a high precision of $\sim 10^{10}$ cm$^{-2}$.[26] As the formation of trions obey the law of mass action involving neutral excitons and free charge carriers (Fig. 3a),[52] the fraction of trions is given as a function of $n$.[26] In addition, the overall PL intensity is sensitive to $n$ because of drastically different nonradiative decay rates of both quasiparticles (Fig. 3b).[15] Based on these facts, we employed in-situ wide-field PL imaging to spatially map the CT process with diffraction-limited spatial and sub-second temporal resolutions (Fig. 3c & 3d).[26] The PL enhancement images obtained by normalizing with respect to a reference image were even more sensitive to $n$ and more revealing because native spatial inhomogeneities could be removed (Fig. 3d). As will be shown below, the method could be coupled to in-situ measurements involving various gaseous and liquid environments.

### 3. Chemical origin of ambient hole doping

In their paper on the first graphene transistor,[11] K. Novoselov et al. briefly reported discovering the spontaneous hole doping in graphene that was supported on silica substrates and exposed to the ambient air. The CT phenomenon was soon verified using Raman spectroscopy[53] and other methods,[54] but its chemical origin and mechanism were not clearly understood. Raman studies on freestanding graphene[13, 55] suggested that it is caused or mediated by underlying silica substrates. However, their role in CT was controversial and claimed to be dopants of electrons[56], holes[57] or both[58] depending on their work function. Interestingly, the CT reaction was greatly enhanced when thermally activated as manifested by the significant upshifts of G peak (Fig. 4a) and change in the charge neutrality point (Fig. 4b),[10, 12] which was confirmed by others but without clarification of its cause.[14, 59-60] Notably, the same spectral changes were attributed to compressive strain,[61-62] which exemplifies the complication of the multimodal sensitivity of Raman probes.

As will be explained below, several experimental breakthroughs led to a conclusion that the spontaneous and activated hole doping phenomena both are caused by ORR and revealed the requirements for the reaction to proceed. In both processes, ORR serves as a half reaction of the redox between $O_2$/$H_2O$ redox couples and hydrated 2-D solids. According to the MG model[28] schematically given in Fig. 1e, the CT direction and rate depend on the energetic alignment and overlap between the frontier levels of both electrochemical entities. For electron (hole) transfer from graphene to the redox couple, the occupied (unoccupied) band of graphene is to be matched with the oxidized (reduced) state of the redox couple. Note that the two states of the couple are displaced by twice of a solvent reorganization energy ($\lambda$). The observed spontaneous electron transfer from graphene to the redox couples indicates that the Fermi level of graphene is closer to the oxidized state of the couple rather than the reduced state. The Nernst equation describing the dependence of the electrochemical potentials on pH and concentration of other involved species provides an experimental method to verify the underlying mechanism, as V. Chakrapani et al.[22] did for the surface conduction of hydrogenated diamond. One crucial question is whether the model and thermodynamic quantities based on bulk aqueous reactions can be applied to the dry systems, which will be discussed below.

The direct link between the CT and $O_2$ was first found for thermally oxidized graphene with nanopores.[12] As shown in Fig. 4c, CT-induced large upshifts in $\omega_G$ could be partially reversed when the samples were exposed to Ar gas instead of air or $O_2$. This fact indicated that some of the hole density is due to weakly bound $O_2$. A similar reversible oxygen sensitivity was observed in partially hydrogenated graphene.[10] A definitive answer was given by an in-situ Raman study with independent control over temperature and gas environment.[13] As shown in Fig. 4d, $\omega_G$ hardly changed upon annealing in Ar at 300 °C but upshifted when graphene was subsequently exposed to $O_2$, which could be reversed in Ar. Remarkably, water vapor itself did not induce but amplified the $O_2$-induced change in $\omega_G$. The spectral changes in the frequency and width of G mode were identical to what was caused by electrical charge injection.[32,33] These facts indicate that annealing modifies the system in a particular manner that favors ORR, which will be explained below. The same paper also found that $O_2$ induces hole carriers in pristine graphene and that $H_2O$ cannot induce but amplifies the reaction. A similar $O_2$-sensitivity was observed from graphene nanoribbons prepared by conventional microfabrication methods.[63] It is to be noted that some of the phonon hardening found in thermally activated graphene originated from compressive strain as quantified by the $\omega_G$-$\omega_{2D}$ correlation analysis (Fig. 4e).[25]

Despite the intimate contact between graphene and the substrates,[64] the lack of doping in suspended graphene samples[13,55] suggested that the CT reaction occurs at the interface, not on the top surface of graphene. We recently found direct evidence using isotopologic bilayers ($^{12}C/^{13}C/SiO_2$), which allowed Raman spectroscopy to depth-profile the charge density (Fig. 5a).[65] This revelation, however, raised a few more questions on the roles of interfacial space and thermal activation, and the efficiency of interfacial molecular diffusion. The clue to the first two questions was provided by the observation that water intercalates into the 2-D interface of pristine[66] and annealed[24] graphene in the presence of excess water (Fig. 5b). This fact implied that the silica surface is hydrophilic enough to induce mass transport through the highly constrained space (Fig. 5c). We verified the conjecture by showing that the degree of the activated CT directly correlates with the hydrophilicity of the silica substrates, measured in terms of water contact angle, in the range of 25 ~ 1000 °C (Fig. 6a & 6b).[26] The thermal activation occurring above 200 °C and deactivation above 800 °C was attributed to the interplay between activated hydroxylation and dehydroxylation of the substrates covered with graphene (Fig. 6c & 6d). Notably, we also showed that it was the silica-bound ambient water that was consumed in the hydroxylation process (Fig. 6e). The last question on diffusion will be answered in the next section.

One critical question was whether the same CT reaction occurs for graphene immersed in oxygenated water. The Nernst equation for ORR predicts that the CT reaction is favored for smaller pH because the oxidized state in Fig. 1e is lowered and overlaps more with the occupied $\pi$ band. As shown in Fig. 7a ~ 7c,[26] graphene was hole-doped in acidic water (pH < 5) containing sufficient $O_2$, and the CT kinetics was faster for lower pH, which is consistent with the electrochemical model. It is also notable that the hole density was in the same order of magnitude as observed for the activated CT (Fig. 6c) and the same pH dependence was confirmed for 2-D semiconductors (Fig. 7d). Based on these facts, we proposed the substrate-assisted ORR model[13, 24-26] to explain the spontaneous and activated CT, where hydrophilic substrates play a pivotal role by accommodating water required for the electrochemical reactions. We also verified that the mechanism is universally operative by showing validity for four kinds of TMDs as will be shown below.[67]

## 4. Interfacial diffusion, CT centers & kinetics

Detection of diffusional progression of CT will be a critical validation of the substrate-assisted ORR because molecular species have only to migrate through the constrained space between the substrates and 2-D materials within the model. Using in-situ time-lapse Raman imaging (Fig. 5b), we observed that the activated CT in graphene was gradually reversed for several hours while the rigid interfacial water layer was formed for the first time.[24] Raman spectroscopy revealed the edge-to-center progression of undoping and dynamic change in lattice strain and structural deformation induced by the intercalating water layers. For the CT process itself that occurs on a time scale of several min,[13] a faster imaging method using PL was adopted.[26] As explained earlier, we established that PL of 2-D semiconducting TMDs serves as a sensitive charge density probe.[26] As the first step, we confirmed that $WS_2$[26] and three other TMDs[67] undergo the spontaneous and reversible CT and become hole-doped in the presence of $O_2$ (Fig. 3b). When supported on hydrophobic hBN, $WS_2$ showed a marginal degree of CT reaction, which agrees with the substrate-assisted ORR.[26] Remarkably, the wide-field PL images in Fig. 8a showed the edge-to-center propagation of PL enhancement after $O_2$ was introduced into the optical gas cell. These real-time snapshots directly captured the interfacial diffusion of $O_2$ (Fig. 8c) in the form of enhanced PL. However, the same sample exhibited a concerted enhancement across the whole basal plane when immersed in water (Fig. 8b), which indicated that CT occurred mainly at the front TMD surfaces (Fig. 8d).[26] This fact verified that the degree of hydration level is a limiting factor and the CT reaction is localized at the TMD-substrate interfaces in a gaseous environment.

The observed diffusion of $O_2$ molecules and their reduction have far-reaching implications because molecular motions and chemical reactions occurring in confined spaces play pivotal roles in catalysis and energy storage applications. It will be interesting if one could control their kinetics by modifying the 2-D vdW space. Indeed, we were able to demonstrate geometric and chemical manipulation of diffusion and CT reactions.[67] As a proof of concept, we first varied the average gap of the vdW space (Fig. 9a). Maximum (~1.6 nm) and intermediate-gap (~0.5 nm) samples could be formed by dry transfer and mechanical exfoliation of TMDs onto $SiO_2$ substrates, respectively (Fig. 9b). Minimum-gap (~0.2 nm) samples were generated by using crystalline hBN or sapphire substrates. As shown in Fig. 9c, larger gap spacing led to more PL contribution from excitons than negative trions in the presence of $O_2$ and, therefore, a higher degree of CT (Fig. 10a). The edge-to-center progression was even faster for the larger gap.[67] This is because the CT rate is determined by the availability and energetics of the composite dopants ($O_2$ and $H_2O$) at the interface. Viewing that the minimum gap is even smaller than the vdW diameter of $O_2$ (0.38 nm), more molecular dopants will be available with increasing the gap because of enhanced diffusion. As schematically shown in Fig. 10b, more extensive hydration cages favored by larger gap spacing will also lower the energy level of the oxidized state, which leads to the increased CT (Fig. 10c).

The chemical nature of the vdW space was also found to govern the diffusion and CT kinetics. TMDs on sapphire exhibited significant interfacial diffusion, unlike hBN-supported samples (Fig. 9d).[67] The drastic contrast originates from the differing hydrophilicity of the two substrates and indicates that the composite dopants can be intercalated into the hydrophilic vdW interface despite the minimum gap. As another chemical control, structural defects were generated on TMD's basal plane that faced the vdW space using UV-generated $O_3$ (Fig. 10f).[67] As shown in Fig. 10e, the overall CT kinetics was substantially accelerated compared to the pristine case. When defects were formed on the opposite basal plane (Fig. 10d), the CT rate was also enhanced without any directional progression.

These observations strongly suggest that the basal-plane defects such as vacancies and antisites serve as the reaction centers for the CT and ensuing ORR.[68] The oxide-based defects may play an additional role by increasing local hydrophilicity. This mechanistic understanding also illuminates the stark kinetic differences observed for four kinds of TMDs as given in Fig. 10g: $MoSe_2 < MoS_2 < WS_2 < WSe_2$ (in the order of increasing CT rate).[67] Because the four vdW spaces are geometrically equivalent owing to their structural similarity, the material dependence is not due to diffusion as verified by the control systems supported on hBN (Fig. 10h), but the energetic alignment between the donor and acceptor states according to the MG model. The leveling and acceleration in their CT rates observed when immersed in water (Fig. 10i) indicate that the material-dependent kinetics is limited by hydration in the four vdW spaces.[67] The differing hydration may originate from the varying density of native defects in the four TMDs.

## 5. Conclusion

In this Account, we have reviewed our experimental endeavor on optical imaging of redox reactions and interfacial molecular diffusion occurring in 2-D crystals supported on dielectric substrates. The original exploration was initiated to elucidate the chemical origin of the elusive hole doping of pristine and thermally activated graphene observed in the ambient air. Using Raman and PL spectroscopies, we developed analytical platforms to quantify charge density in situ and with high precision in graphene and TMDs, respectively. The optical tools allowed us to unravel that the chemical doping is induced by the oxygen reduction reaction involving $O_2$, $H_2O$ and electrons from 2-D crystals. Its electrochemical nature was verified by its dependence on $O_2$, pH and water based on the Marcus-Gerischer model and led to the localization of CT at the interface between 2-D crystals and hydrophilic substrates. Wide-field optical imaging also revealed that the interface serves as a route for molecular transport required for the CT reactions. Geometric and chemical modification of the interfacial gap space led to drastic changes in the diffusion and CT kinetics, respectively. Our findings on the mechanism of the fundamental charge exchange will be essential in exploring material and device properties of low dimensional materials.

# AUTHOR INFORMATION

**Corresponding author**

*E-mail: sunryu@postech.ac.kr

[#]These authors contributed equally

**Notes**

The authors declare no competing financial interest.

**Biographies**

**Haneul Kang** received her B.S. in Department of Chemistry at Ajou University in 2015 and her Ph.D. in Department of Chemistry at POSTECH in 2021. She is currently a postdoctoral researcher at POSTECH.

**Kwanghee Park** received his B.S. in Department of Applied Chemistry at Kyunghee University in 2015 and his Ph.D. in Department of Chemistry at POSTECH in 2021. He is currently a postdoctoral researcher at Korean Research Institute of Standards and Science (KRISS).

**Sunmin Ryu** is an associate professor in Department of Chemistry at POSTECH. He obtained his B.S. (1998), M.S. (2000), and Ph.D. (2005) in Department of Chemistry at Seoul National University. He worked as a postdoctoral researcher at Korea Research Institute of Standards and Science (KRISS) until 2006 and then at the chemistry department of Columbia University until 2009. He started his independent research and teaching in Department of Applied Chemistry at Kyung Hee University as an assistant professor and moved to POSTECH in 2014. His research interests include energy and charge transfer dynamics, photophysics and photochemistry of low-dimensional materials.

# ACKNOWLEDGEMENT

S.R. acknowledges the financial support from Samsung Research Funding Center of Samsung Electronics under Project Number SSTF-BA1702-08.

# REFERENCES

1. Ferraris, J.; Cowan, D. O.; Walatka, V.; Perlstein, J. H. Electron transfer in a new highly conducting donor-acceptor complex. *J. Am. Chem. Soc.* **1973,** *95* (3), 948-949.
2. Ishii, H.; Sugiyama, K.; Ito, E.; Seki, K. Energy level alignment and interfacial electronic structures at organic metal and organic organic interfaces. *Adv. Mater.* **1999,** *11* (8), 605-625.
3. Chiang, C. K.; Fincher, C. R.; Park, Y. W.; Heeger, A. J.; Shirakawa, H.; Louis, E. J.; Gau, S. C.; Macdiarmid, A. G. Electrical-conductivity in doped polyacetylene. *Phys. Rev. Lett.* **1977,** *39* (17), 1098-1101.
4. Grigorian, L.; Williams, K. A.; Fang, S.; Sumanasekera, G. U.; Loper, A. L.; Dickey, E. C.; Pennycook, S. J.; Eklund, P. C. Reversible intercalation of charged iodine chains into carbon nanotube ropes. *Phys. Rev. Lett.* **1998,** *80* (25), 5560-5563.
5. Lee, R. S.; Kim, H. J.; Fischer, J. E.; Thess, A.; Smalley, R. E. Conductivity enhancement in single-walled carbon nanotube bundles doped with K and Br. *Nature* **1997,** *388* (6639), 255-257.
6. Ohta, T.; Bostwick, A.; Seyller, T.; Horn, K.; Rotenberg, E. Controlling the electronic structure of bilayer graphene. *Science* **2006,** *313* (5789), 951-954.
7. Jung, N.; Kim, N.; Jockusch, S.; Turro, N. J.; Kim, P.; Brus, L. Charge transfer chemical doping of few layer graphenes: charge distribution and band gap formation. *Nano Lett.* **2009,** *9* (12), 4133–4137.
8. Jung, N.; Kim, B.; Crowther, A. C.; Kim, N.; Nuckolls, C.; Brus, L. Optical reflectivity and Raman scattering in few-layer-thick graphene highly doped by K and Rb. *ACS Nano* **2011,** *5* (7), 5708-5716.
9. Wang, Y.; Zheng, Y.; Han, C.; Chen, W. Surface charge transfer doping for two-dimensional semiconductor-based electronic and optoelectronic devices. *Nano Research* **2021,** *14* (6), 1682-1697.
10. Ryu, S.; Han, M. Y.; Maultzsch, J.; Heinz, T. F.; Kim, P.; Steigerwald, M. L.; Brus, L. E. Reversible basal plane hydrogenation of graphene. *Nano Lett.* **2008,** *8* (12), 4597-4602.
11. Novoselov, K. S.; Geim, A. K.; Morozov, S. V.; Jiang, D.; Zhang, Y.; Dubonos, S. V.; Grigorieva, I. V.; Firsov, A. A. Electric field effect in atomically thin carbon films. *Science* **2004,** *306* (5696), 666-669.
12. Liu, L.; Ryu, S.; Tomasik, M. R.; Stolyarova, E.; Jung, N.; Hybertsen, M. S.; Steigerwald, M. L.; Brus, L. E.; Flynn, G. W. Graphene oxidation: thickness dependent etching and strong chemical doping. *Nano Lett.* **2008,** *8*, 1965-1970.
13. Ryu, S.; Liu, L.; Berciaud, S.; Yu, Y.-J.; Liu, H.; Kim, P.; Flynn, G. W.; Brus, L. E. Atmospheric oxygen binding and hole doping in deformed graphene on a $SiO_2$ substrate. *Nano Lett.* **2010,** *10* (12), 4944-4951.
14. Abdula, D.; Ozel, T.; Kang, K.; Cahill, D. G.; Shim, M. Environment-induced effects on the temperature dependence of raman spectra of single-layer graphene. *J. Phys. Chem. C* **2008,** *112* (51), 20131-20134.
15. Tongay, S.; Zhou, J.; Ataca, C.; Liu, J.; Kang, J. S.; Matthews, T. S.; You, L.; Li, J.; Grossman, J. C.; Wu, J. Broad-range modulation of light emission in two-dimensional semiconductors by molecular physisorption gating. *Nano Lett.* **2013,** *13* (6), 2831-2836.
16. Nan, H.; Wang, Z.; Wang, W.; Liang, Z.; Lu, Y.; Chen, Q.; He, D.; Tan, P.; Miao, F.; Wang, X.; Wang, J.; Ni, Z. Strong photoluminescence enhancement of $MoS_2$ through defect engineering and oxygen bonding. *ACS Nano* **2014,** *8* (6), 5738-5745.
17. Lekawa-Raus, A.; Kurzepa, L.; Kozlowski, G.; Hopkins, S. C.; Wozniak, M.; Lukawski, D.;

Glowacki, B. A.; Koziol, K. K. Influence of atmospheric water vapour on electrical performance of carbon nanotube fibres. *Carbon* **2015,** *87*, 18-28.
18. Collins, P. G.; Bradley, K.; Ishigami, M.; Zettl, A. Extreme oxygen sensitivity of electronic properties of carbon nanotubes. *Science* **2000,** *287*, 1801.
19. Derycke, V.; Martel, R.; Appenzeller, J.; Avouris, P. Controlling doping and carrier injection in carbon nanotube transistors. *Appl. Phys. Lett.* **2002,** *80*, 2773.
20. Gaur, A.; Shim, M. Substrate-enhanced $O_2$ adsorption and complexity in the Raman G-band spectra of individual metallic carbon nanotubes. *Phys. Rev. B* **2008,** *78* (12), 125422.
21. Landstrass, M. I.; Ravi, K. V. Hydrogen passivation of electrically active defects in diamond. *Appl. Phys. Lett.* **1989,** *55* (14), 1391-1393.
22. Chakrapani, V.; Angus, J. C.; Anderson, A. B.; Wolter, S. D.; Stoner, B. R.; Sumanasekera, G. U. Charge transfer equilibria between diamond and an aqueous oxygen electrochemical redox couple. *Science* **2007,** *318*, 1424-1430.
23. Levesque, P. L.; Sabri, S. S.; Aguirre, C. M.; Guillemette, J.; Siaj, M.; Desjardins, P.; Szkopek, T.; Martel, R. Probing charge transfer at surfaces using graphene transistors. *Nano Lett.* **2011,** *11* (1), 132-137.
24. Lee, D.; Ahn, G.; Ryu, S. Two-dimensional water diffusion at a graphene-silica interface. *J. Am. Chem. Soc.* **2014,** *136* (18), 6634-6642.
25. Lee, J. E.; Ahn, G.; Shim, J.; Lee, Y. S.; Ryu, S. Optical separation of mechanical strain from charge doping in graphene. *Nat. Commun.* **2012,** *3*, 1024.
26. Park, K.; Kang, H.; Koo, S.; Lee, D.; Ryu, S. Redox-governed charge doping dictated by interfacial diffusion in two-dimensional materials. *Nat. Commun.* **2019,** *10* (1), 4931.
27. Maier, F.; Riedel, M.; Mantel, B.; Ristein, J.; Ley, L. Origin of surface conductivity in diamond. *Phys. Rev. Lett.* **2000,** *85* (16), 3472-3475.
28. Memming, R. *Semiconductor electrochemistry*. Wiley-VCH Verlag: 2001.
29. Kim, W.; Javey, A.; Vermesh, O.; Wang, Q.; Li, Y.; Dai, H. Hysteresis caused by water molecules in carbon nanotube field-effect transistors. *Nano Lett.* **2003,** *3* (2), 193-198.
30. Aguirre, C. M.; Levesque, P. L.; Paillet, M.; Lapointe, F.; St-Antoine, B. C.; Desjardins, P.; Martel, R. The role of the oxygen/water redox couple in suppressing electron conduction in field-effect transistors. *Adv. Mater.* **2009,** *21* (30), 3087-3091.
31. Crowther, A. C.; Ghassaei, A.; Jung, N.; Brus, L. E. Strong charge-transfer doping of 1 to 10 layer graphene by $NO_2$. *ACS Nano* **2012,** *6* (2), 1865-1875.
32. Yan, J.; Zhang, Y.; Kim, P.; Pinczuk, A. Electric field effect tuning of electron-phonon coupling in graphene. *Phys. Rev. Lett.* **2007,** *98*, 166802/1-166802/4.
33. Pisana, S.; Lazzeri, M.; Casiraghi, C.; Novoselov, K. S.; Geim, A. K.; Ferrari, A. C.; Mauri, F. Breakdown of the adiabatic Born-Oppenheimer approximation in graphene. *Nat. Mater.* **2007,** *6* (3), 198-201.
34. Mohiuddin, T. M. G.; Lombardo, A.; Nair, R. R.; Bonetti, A.; Savini, G.; Jalil, R.; Bonini, N.; Basko, D. M.; Galiotis, C.; Marzari, N.; Novoselov, K. S.; Geim, A. K.; Ferrari, A. C. Uniaxial strain in graphene by Raman spectroscopy: G peak splitting, Grüneisen parameters, and sample orientation. *Phys. Rev. B* **2009,** *79*, 205433.
35. Cançado, L. G.; Jorio, A.; Ferreira, E. H. M.; Stavale, F.; Achete, C. A.; Capaz, R. B.; Moutinho, M. V. O.; Lombardo, A.; Kulmala, T. S.; Ferrari, A. C. Quantifying defects in graphene via Raman spectroscopy at different excitation energies. *Nano Lett.* **2011,** *11* (8), 3190-3196.
36. Ferrari, A. C.; Basko, D. M. Raman spectroscopy as a versatile tool for studying the properties of graphene. *Nat. Nanotechnol.* **2013,** *8* (4), 235-246.
37. Lee, C.; Yan, H.; Brus, L. E.; Heinz, T. F.; Hone, J.; Ryu, S. Anomalous lattice vibrations of single- and few-Layer $MoS_2$. *ACS Nano* **2010,** *4* (5), 2695-2700.

38. Stolyarova, E.; Rim, K. T.; Ryu, S.; Maultzsch, J.; Kim, P.; Brus, L. E.; Heinz, T. F.; Hybertsen, M. S.; Flynn, G. W. High-resolution scanning tunneling microscopy imaging of mesoscopic graphene sheets on an insulating surface. *Proc. Natl. Acad. Sci. U. S. A.* **2007,** *104* (22), 9209-9212.
39. Das, A.; Chakraborty, B.; Piscanec, S.; Pisana, S.; Sood, A. K.; Ferrari, A. C. Phonon renormalization in doped bilayer graphene. *Phys. Rev. B* **2009,** *79* (15), 155417.
40. Yoon, D.; Son, Y. W.; Cheong, H. Strain-dependent splitting of the double-resonance Raman scattering band in graphene. *Phys. Rev. Lett.* **2011,** *106* (15), 155502.
41. Hwang, C.; Siegel, D. A.; Mo, S.-K.; Regan, W.; Ismach, A.; Zhang, Y.; Zettl, A.; Lanzara, A. Fermi velocity engineering in graphene by substrate modification. *Sci. Rep.* **2012,** *2*, 590.
42. Ferrari, A. C.; Meyer, J. C.; Scardaci, V.; Casiraghi, C.; Lazzeri, M.; Mauri, F.; Piscanec, S.; Jiang, D.; Novoselov, K. S.; Roth, S.; Geim, A. K. Raman spectrum of graphene and graphene layers. *Phys. Rev. Lett.* **2006,** *97* (18), 187401/1-187401/4.
43. Ni, Z. H.; Wang, Y. Y.; Yu, T.; You, Y. M.; Shen, Z. X. Reduction of Fermi velocity in folded graphene observed by resonance Raman spectroscopy. *Phys. Rev. B* **2008,** *77* (23), 235403.
44. Ahn, G.; Kim, H. R.; Ko, T. Y.; Choi, K.; Watanabe, K.; Taniguchi, T.; Hong, B. H.; Ryu, S. Optical probing of the electronic interaction between graphene and hexagonal boron nitride. *ACS Nano* **2013,** *7* (2), 1533-1541.
45. Choi, J.; Koo, S.; Song, M.; Jung, D. Y.; Choi, S.-Y.; Ryu, S. Varying electronic coupling at graphene–copper interfaces probed with Raman spectroscopy. *2D Mater.* **2020,** *7* (2), 025006.
46. Mafra, D. L.; Samsonidze, G.; Malard, L. M.; Elias, D. C.; Brant, J. C.; Plentz, F.; Alves, E. S.; Pimenta, M. A. Determination of LA and TO phonon dispersion relations of graphene near the Dirac point by double resonance Raman scattering. *Phys. Rev. B* **2007,** *76* (23), 233407.
47. Ahn, G.; Ryu, S. Reversible sulfuric acid doping of graphene probed by in-situ multi-wavelength Raman spectroscopy. *Carbon* **2018,** *138*, 257-263.
48. Kim, S.; Ryu, S. Thickness-dependent native strain in graphene membranes visualized by Raman spectroscopy. *Carbon* **2016,** *100*, 283-290.
49. Chakraborty, B.; Bera, A.; Muthu, D. V. S.; Bhowmick, S.; Waghmare, U. V.; Sood, A. K. Symmetry-dependent phonon renormalization in monolayer $MoS_2$ transistor. *Phys. Rev. B* **2012,** *85* (16), 161403.
50. Rice, C.; Young, R. J.; Zan, R.; Bangert, U.; Wolverson, D.; Georgiou, T.; Jalil, R.; Novoselov, K. S. Raman-scattering measurements and first-principles calculations of strain-induced phonon shifts in monolayer $MoS_2$. *Phys. Rev. B* **2013,** *87* (8), 081307.
51. Mak, K. F.; He, K.; Lee, C.; Lee, G. H.; Hone, J.; Heinz, T. F.; Shan, J. Tightly bound trions in monolayer $MoS_2$. *Nat. Mater.* **2013,** *12* (3), 207-211.
52. Siviniant, J.; Scalbert, D.; Kavokin, A. V.; Coquillat, D.; Lascaray, J. P. Chemical equilibrium between excitons, electrons, and negatively charged excitons in semiconductor quantum wells. *Phys. Rev. B* **1999,** *59* (3), 1602-1604.
53. Casiraghi, C.; Pisana, S.; Novoselov, K. S.; Geim, A. K.; Ferrari, A. C. Raman fingerprint of charged impurities in graphene. *Appl. Phys. Lett.* **2007,** *91*, 233108/1-233108/3.
54. Dean, C. R.; Young, A. F.; Meric, I.; Lee, C.; Wang, L.; Sorgenfrei, S.; Watanabe, K.; Taniguchi, T.; Kim, P.; Shepard, K. L.; Hone, J. Boron nitride substrates for high-quality graphene electronics. *Nat. Nanotechnol.* **2010,** *5* (10), 722-726.
55. Berciaud, S.; Ryu, S.; Brus, L. E.; Heinz, T. F. Probing the intrinsic properties of exfoliated graphene: Raman spectroscopy of free-standing monolayers. *Nano Lett.* **2009,** *9* (1), 346-352.
56. Romero, H. E.; Shen, N.; Joshi, P.; Gutierrez, H. R.; Tadigadapa, S. A.; Sofo, J. O.; Eklund, P. C. n-type behavior of graphene supported on $Si/SiO_2$ substrates. *ACS Nano* **2008,** *2*, 2037-2044.
57. Hossain, M. Z. Chemistry at the graphene-$SiO_2$ interface. *Appl. Phys. Lett.* **2009,** *95* (14),

143125.
58. Shi, Y. M.; Dong, X. C.; Chen, P.; Wang, J. L.; Li, L. J. Effective doping of single-layer graphene from underlying SiO$_2$ substrates. *Phys. Rev. B* **2009,** *79* (11), 115402.
59. Ni, Z. H.; Wang, H. M.; Luo, Z. Q.; Wang, Y. Y.; Yu, T.; Wu, Y. H.; Shen, Z. X. The effect of vacuum annealing on graphene. *J. Raman Spectrosc.* **2010,** *41* (5), 479-483.
60. Nourbakhsh, A.; Cantoro, M.; Klekachev, A.; Clemente, F.; Soree, B.; van der Veen, M. H.; Vosch, T.; Stesmans, A.; Sels, B.; De Gendt, S. Tuning the Fermi level of SiO$_2$-supported single-layer graphene by thermal annealing. *J. Phys. Chem. C* **2010,** *114* (15), 6894-6900.
61. Ni, Z. H.; Wang, H. M.; Ma, Y.; Kasim, J.; Wu, Y. H.; Shen, Z. X. Tunable stress and controlled thickness modification in graphene by annealing. *ACS Nano* **2008,** *2* (5), 1033-1039.
62. Chen, C.-C.; Bao, W.; Theiss, J.; Dames, C.; Lau, C. N.; Cronin, S. B. Raman spectroscopy of ripple formation in suspended graphene. *Nano Lett.* **2009,** *9* (12), 4172-4176.
63. Ryu, S.; Maultzsch, J.; Han, M. Y.; Kim, P.; Brus, L. E. Raman spectroscopy of lithographically patterned graphene nanoribbons. *ACS Nano* **2011,** *5* (5), 4123-4130.
64. Bunch, J. S.; Verbridge, S. S.; Alden, J. S.; Zande, A. M. v. d.; Parpia, J. M.; Craighead, H. G.; McEuen, P. L. Impermeable atomic membranes from graphene sheets. *Nano Lett.* **2008,** *8*, 2458-2462.
65. Jeon, H.; Teraji, T.; Watanabe, K.; Taniguchi, T.; Ryu, S. Lattice vibrations of single and multi-layer isotopologic graphene. *Carbon* **2018,** *140*, 449-457.
66. Lee, M. J.; Choi, J. S.; Kim, J. S.; Byun, I. S.; Lee, D. H.; Ryu, S.; Lee, C.; Park, B. H. Characteristics and effects of diffused water between graphene and a SiO$_2$ substrate. *Nano Research* **2012,** *5* (10), 710-717.
67. Kang, H.; Ryu, S. Optical imaging of chemically and geometrically controlled molecular diffusion and redox in 2D van der Waals space. *J. Phys. Chem. C* **2021,** *ASAP*.
68. Guo, D.; Shibuya, R.; Akiba, C.; Saji, S.; Kondo, T.; Nakamura, J. Active sites of nitrogen-doped carbon materials for oxygen reduction reaction clarified using model catalysts. *Science* **2016,** *351* (6271), 361-365.

**Figure & Captions**

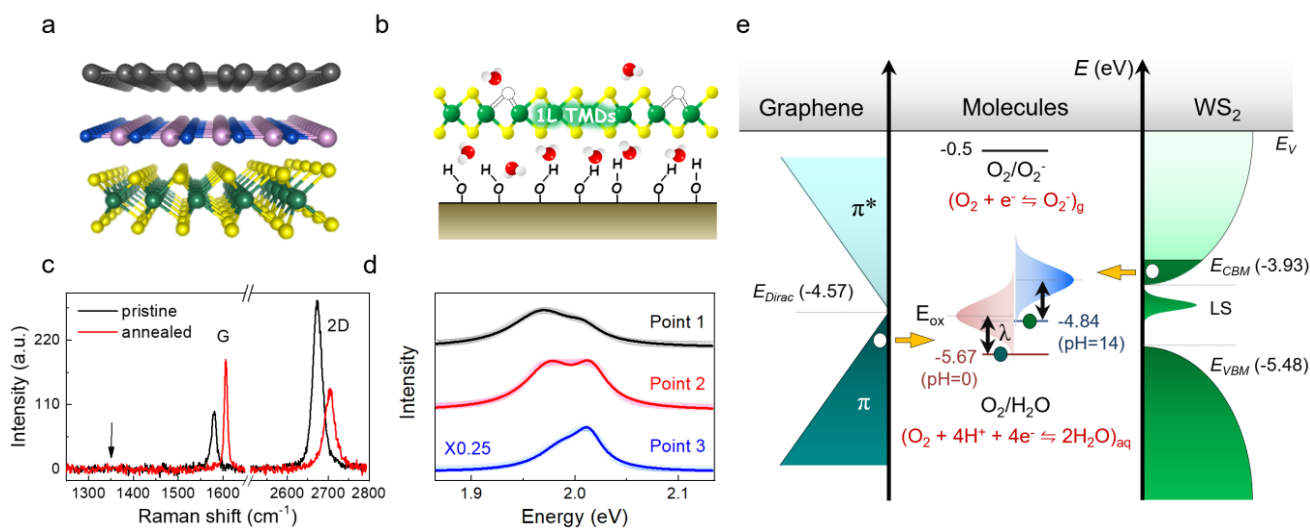

**Figure 1. CT doping by oxygen/water redox couple.** (a) Structures of 1L graphene, hBN and TMD (top to bottom). (b) Side-view scheme for 1L TMD supported on hydrophilic substrate with water molecules. (c) Representative Raman spectra of graphene obtained before (black) and after (red) thermal annealing at 400 °C for 2 h. (d) PL spectra of pristine 1L $WS_2/SiO_2$ obtained from different positions. (e) Energy level diagram for hole doping of graphene and $WS_2$ by $O_2/H_2O$ redox couples. Adapted with permission from ref. 25 (Copyright 2012 Springer Nature) and ref. 26 (Copyright 2019 Springer Nature)

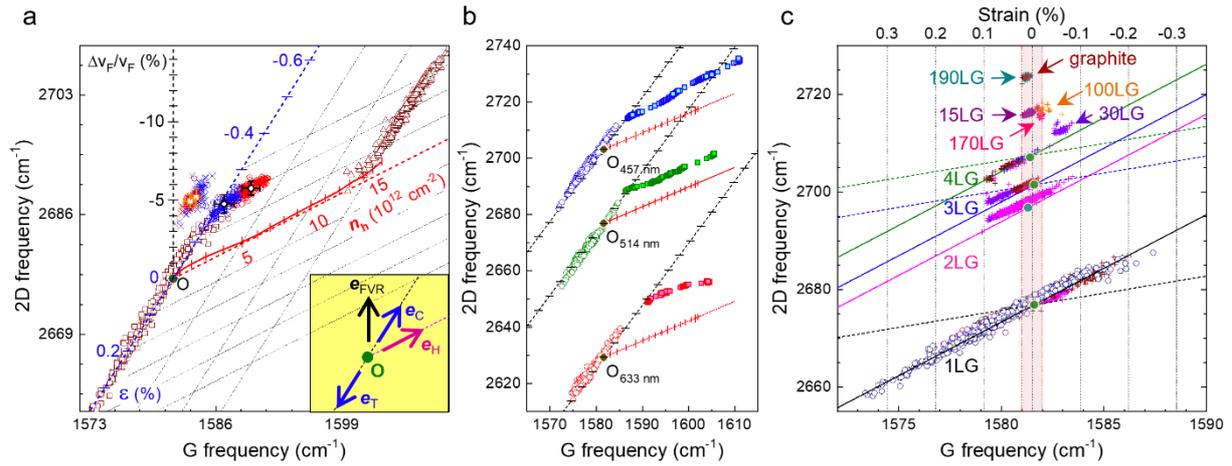

**Figure 2. Raman metrology for charge density.** (a) Correlation between $\omega_G$ and $\omega_{2D}$ of graphene on hBN (crosses) and SiO$_2$ (circles), respectively. Inset: Arrows labeled **e$_T$**, **e$_C$**, **e$_H$**, and **e$_{FVR}$** represent the trajectories of O($\omega_G$, $\omega_{2D}$) affected, respectively, by tensile strain, compressive strain, hole doping and vdW interlayer interaction leading to Fermi velocity reduction (FVR). (b) Strain-charge density analysis for 457, 514 and 633 nm. (c) Strain-dominated Raman spectroscopic variations in graphene samples of various thickness. Adapted with permission from ref. 44 (Copyright 2013 American Chemical Society), ref. 47 (Copyright 2018 Elsevier) and ref. 48 (Copyright 2016 Elsevier).

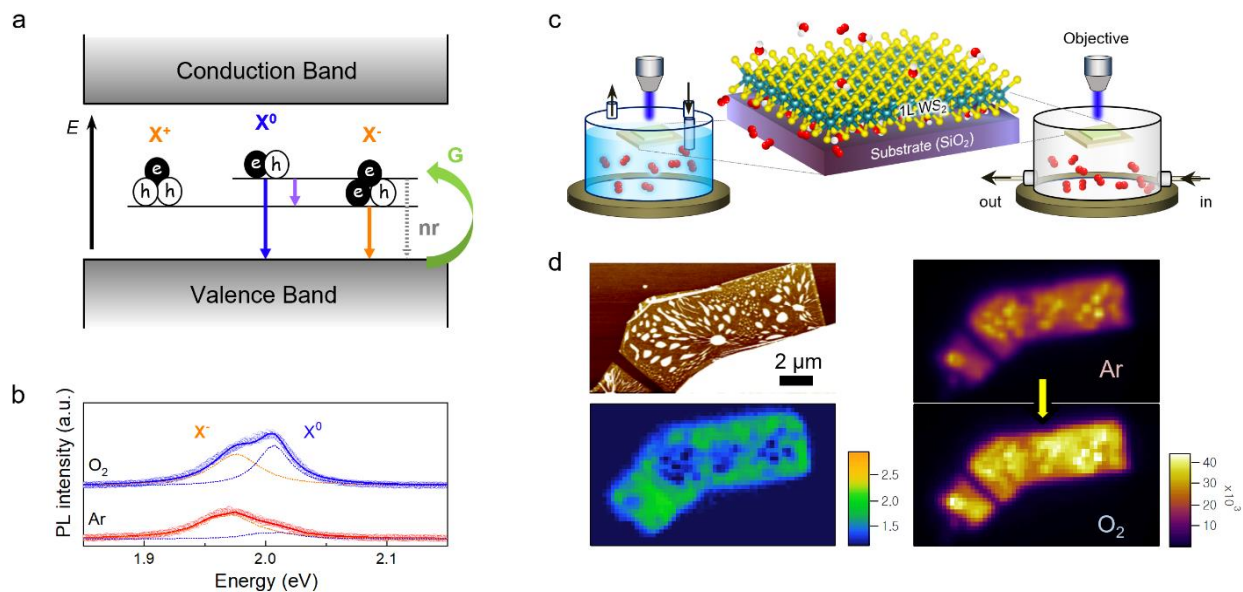

**Figure 3. Photoluminescence metrology for charge density.** (a) Photoexcitation of neutral ($X^0$) and charged ($X^+$ and $X^-$) excitonic states and their recombination processes. (b) Charge-density-dependent PL spectra of 1L $WS_2$ affected by gas environment. (c) In-situ PL imaging in $O_2$-controlled gaseous (right) or aqueous (left) environment. (d) AFM topographic (top left) and PL enhancement (bottom left) images of 1L $WS_2$. The enhancement was defined as PL signal in $O_2$ (bottom right) divided by that in Ar (top right). Adapted with permission from ref. 26 (Copyright 2019 Springer Nature).

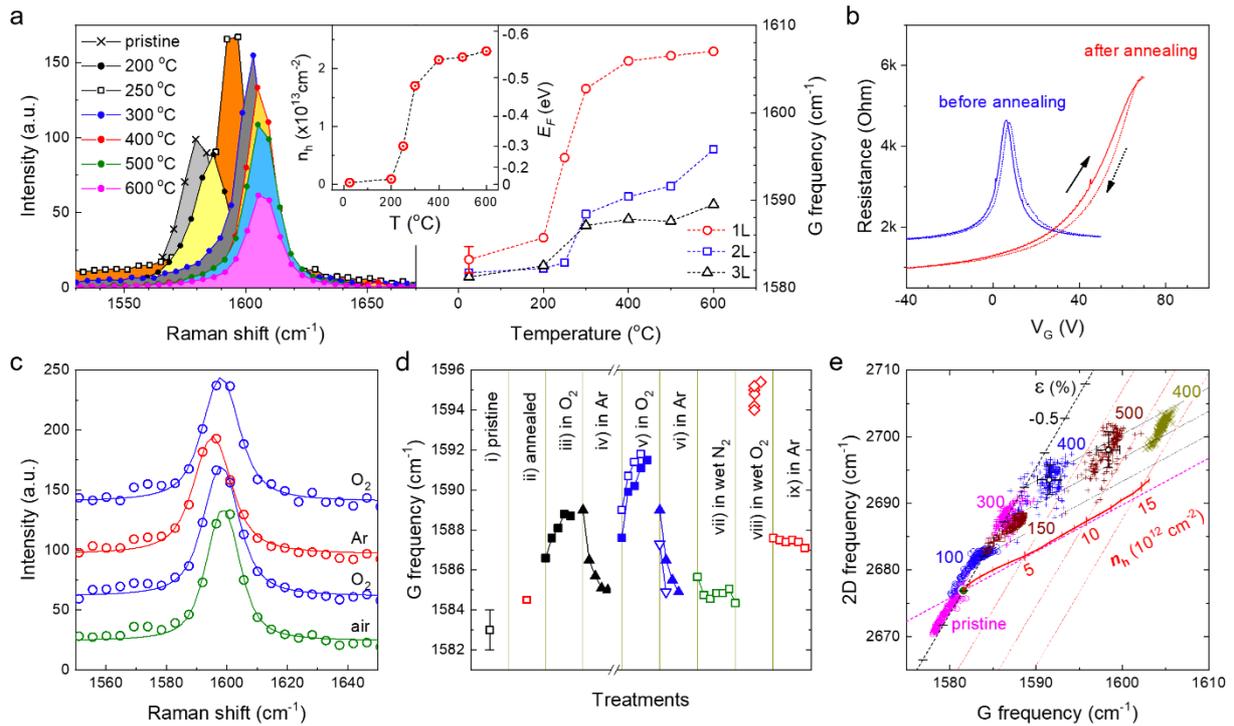

**Figure 4. Chemical origin of ambient hole doping.** (a) G Raman spectra of oxidized 1L graphene (left) and $\omega_G$ of 1 ~ 3L (right) given as a function of oxidation temperature. Inset: $n_h$ of oxidized 1L graphene. (b) Electrical resistance of 1L graphene device in the air as a function of back-gate bias ($V_G$): before and after 2 h annealing at 350 °C in Ar atmosphere. (c) G Raman spectra of oxidized 1L graphene in various gases. (d) Effects of various gases on $\omega_G$ of 1L graphene before and after annealing at 290 °C. (e) Thermally induced variations in ($\omega_G$, $\omega_{2D}$) by successive annealing cycles. Adapted with permission from ref. 12 (Copyright 2008 American Chemical Society), ref. 13 (Copyright 2010 American Chemical Society) and ref. 25 (Copyright 2012 Springer Nature).

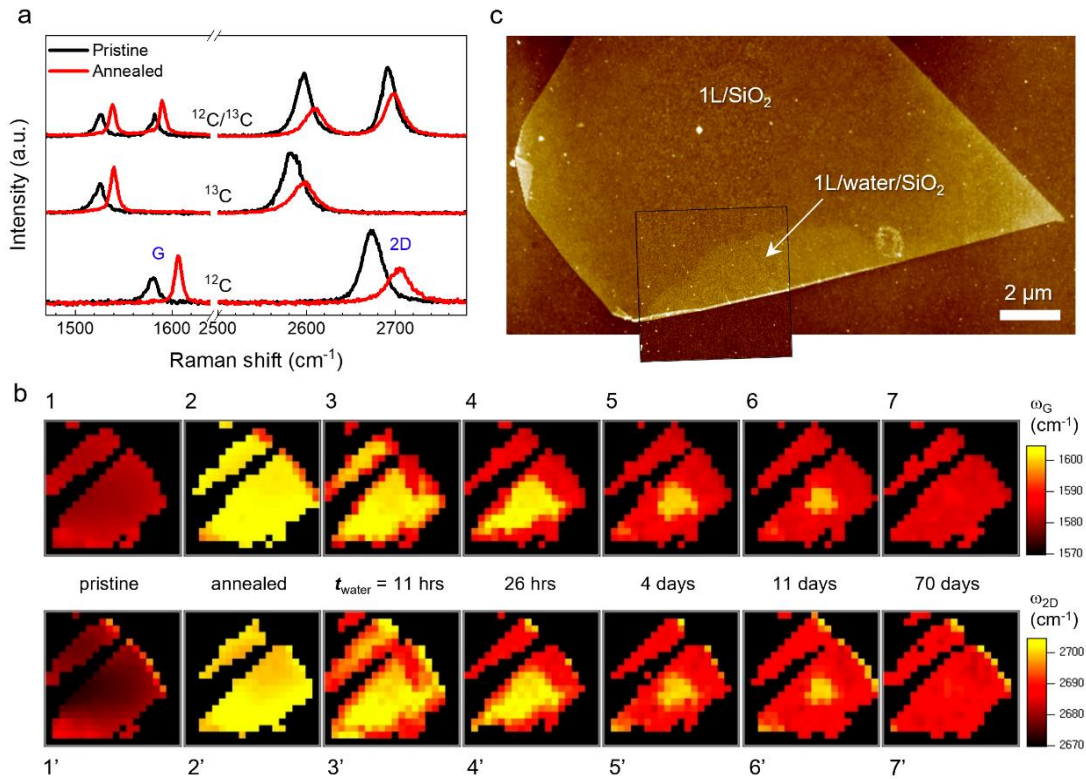

**Figure 5. Interfacial nature of hole doping.** (a) Raman spectra of $^{12}C/^{13}C$ 2L, $^{13}C$ 1L and $^{12}C$ 1L graphene obtained before and after annealing at 400 ºC. (b) Raman maps of $\omega_G$ (1 ~ 7) and $\omega_{2D}$ (1' ~ 7') obtained for 1L graphene in its pristine, annealed and submerged states for various submersion time ($t_{water}$) in water. Each image is 15 x 15 μm$^2$. (c) Noncontact AFM height images of annealed 1L graphene obtained after submersion in water for 14 h. The plateau in the lower portion of the sample is due to an interfacial water layer. Adapted with permission from ref. 65 (Copyright 2018 Elsevier) and ref. 24 (Copyright 2014 American Chemical Society).

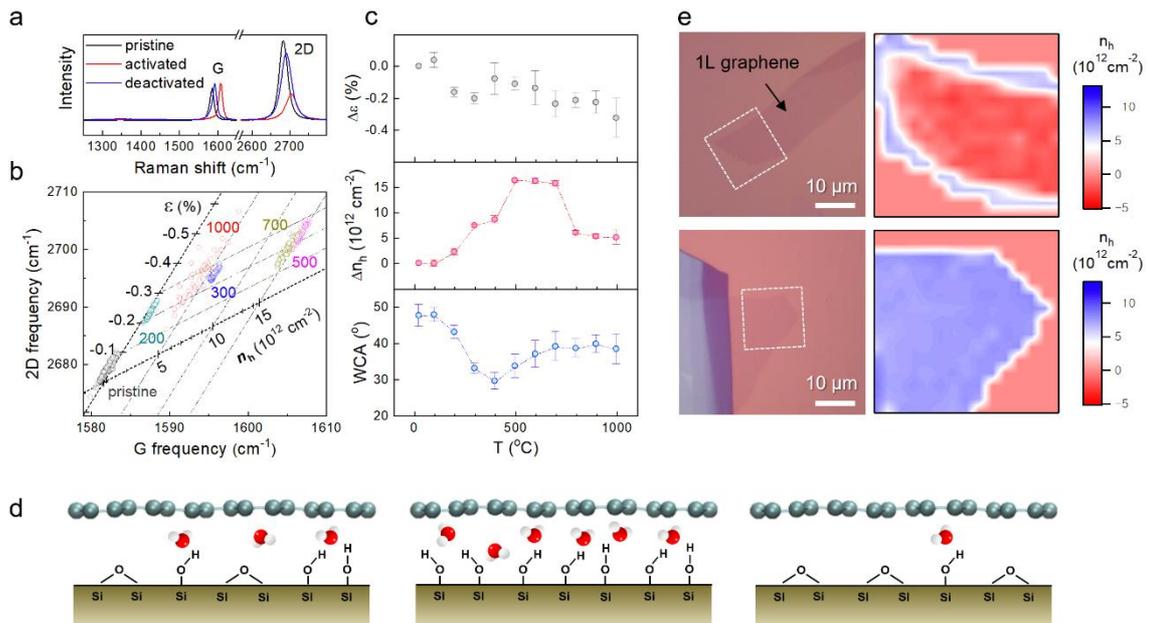

**Figure 6. Mechanism of thermal activation.** (a) Raman spectra of pristine, thermally activated (annealed at 500 °C), and deactivated (annealed at 1000 °C) 1L graphene. (b) Correlation between $\omega_G$ and $\omega_{2D}$ of 1L graphene annealed at 200 ~ 1000 °C. (c) Changes in lattice strain (top) and hole-density (middle) of graphene; water-contact angle (bottom) of $SiO_2$ substrates as a function of annealing temperature. (d) Schematic side views of air-equilibrated graphene–$SiO_2$ interface with pristine (left), hydroxylated by thermal activation (middle) and dehydroxylated substrates by thermal deactivation (right). (e) Optical micrographs (left) and $n_h$ images (right) of annealed 1L graphene samples prepared on diethyl zinc-treated (top) and non-treated (bottom) $SiO_2$ substrates. Adapted with permission from ref. 26 (Copyright 2019 Springer Nature).

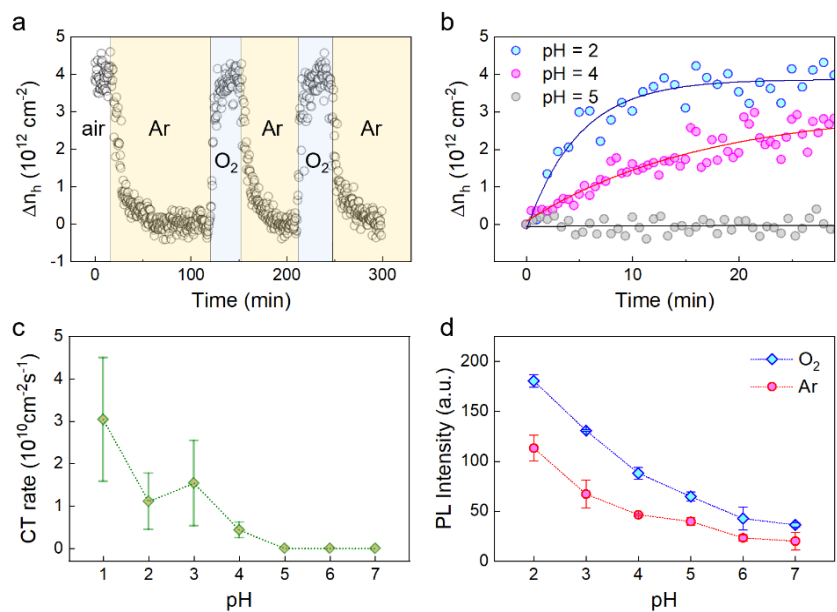

**Figure 7. Electrochemical nature.** (a) Time-lapse measurements of $n_h$ in 1L graphene in HCl solution (pH = 2) through which Ar and $O_2$ gases were sparged alternatively. (b) pH-dependent kinetics of $O_2$-induced rise in $n_h$ of 1L graphene. (c) Initial CT rate per unit area of 1L graphene in HCl solution of various pH. (d) PL intensity of 1L $WS_2$ modulated by $O_2$ dissolved in HCl solution of various pH. Adapted with permission from ref. 26 (Copyright 2019 Springer Nature).

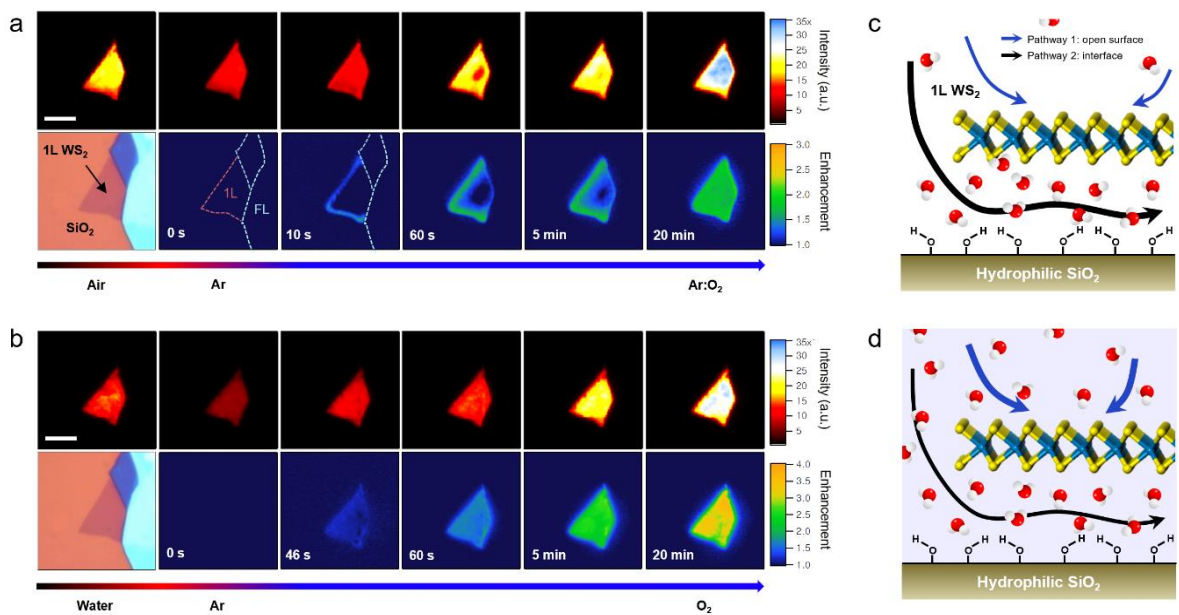

**Figure 8. Real-time photoluminescence images of interface-confined redox reactions.** (a & b) Time-lapse PL (top rows) and enhancement (bottom rows) images of 1L $WS_2$ in gaseous (a) and aqueous (b) environments obtained with increasing $O_2$ concentration. Scale bars are 8 μm. (c & d) Schemes for major CT routes of 1L $WS_2$ in gaseous (c) and aqueous (d) environments. Adapted with permission from ref. 26 (Copyright 2019 Springer Nature).

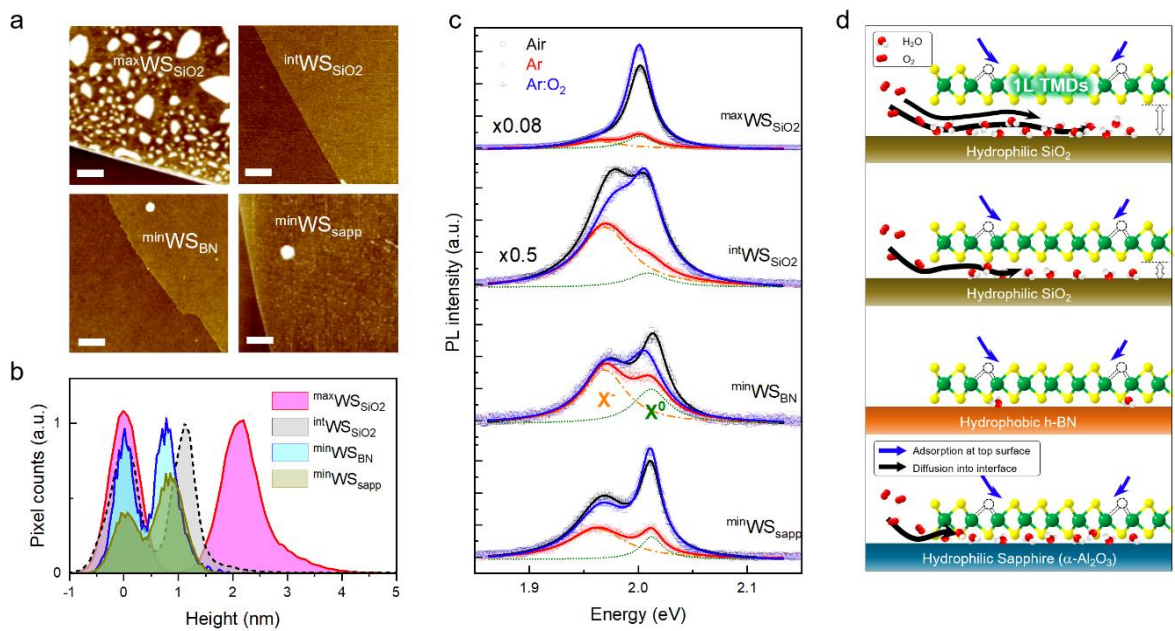

**Figure 9. Geometric control of interfacial CT reaction.** (a & b) AFM topographic images (a) and height histograms (b) of four $WS_2$ samples with various interfacial gaps against different substrates. Scale bars are 0.5 μm. (c) Effects of $O_2$ on PL spectra of the four $WS_2$ samples. (d) Side-view schemes for interfacial diffusion (black arrows) and surface adsorption (blue arrows) of $O_2$ in the four $WS_2$ systems. Adapted with permission from ref. 67 (Copyright 2021 American Chemical Society).

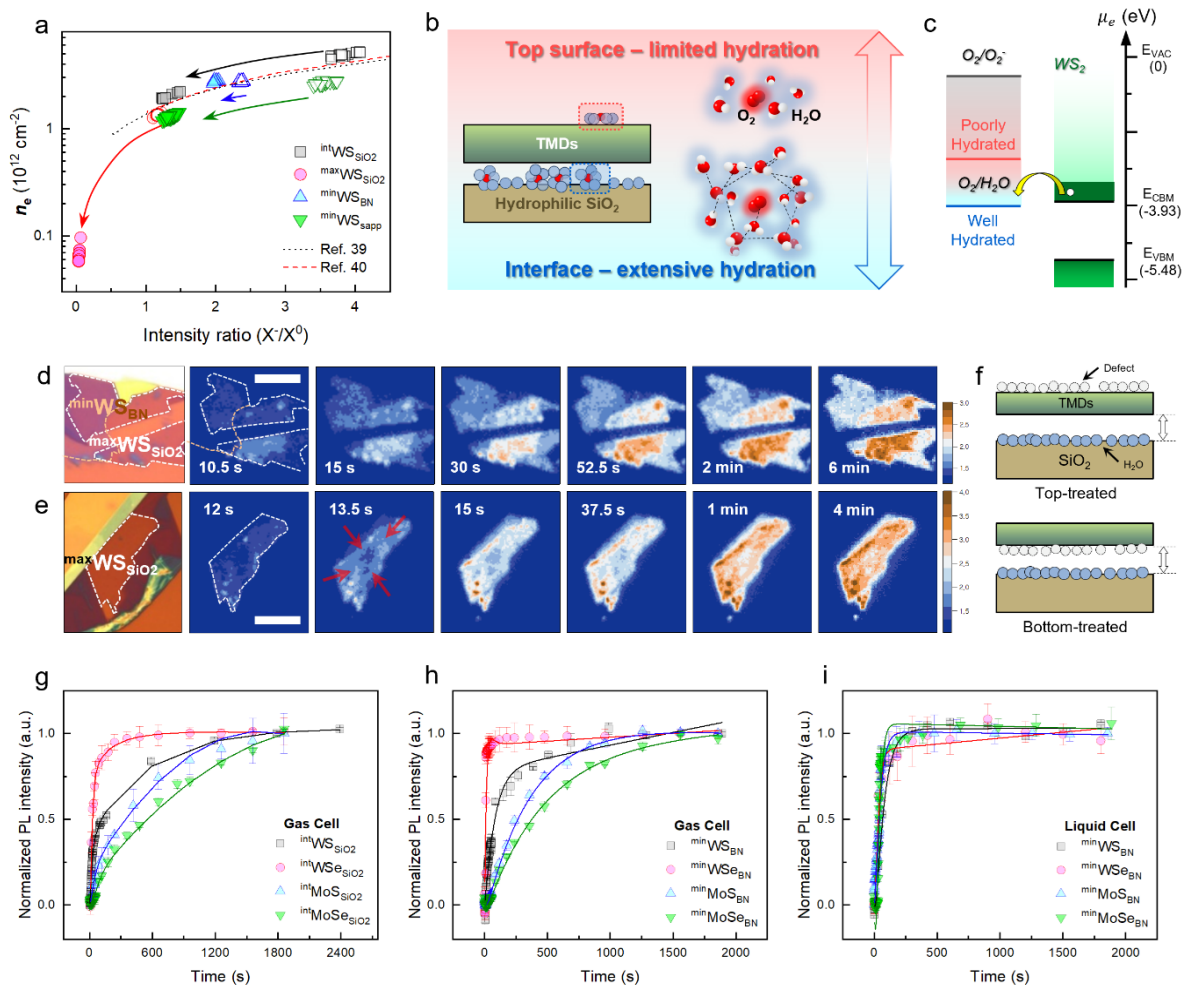

**Figure 10. Charge transfer reaction promoted by chemical control.** (a) Electron density ($n_e$) in Ar (open) and $O_2$ (filled) given as a function of PL intensity ratio ($X^-/X^0$). (b) Scheme illustrating good and poor hydration at hydrophilic interface and hydrophobic TMD surface, respectively. (c) Energy-level diagram for CT from $WS_2$ to $O_2/H_2O$ redox couple. (d & e) Optical micrographs (first column) and time-lapse PL enhancement images of $WS_2$: top (d) and bottom (e) surfaces of $WS_2$ were ozone-treated to induce defects, respectively. Scale bars are 7 (d) and 10 (e) μm. (f) Schematic side views of top- (d) and bottom- (e) treated samples. (g ~ i) Normalized PL enhancement of 4 different 1L TMDs given as a function of $O_2$-exposure time: TMD/$SiO_2$ in gaseous environment (g), TMD/hBN in gaseous environment (h), and TMD/hBN in distilled water (i). Adapted with permission from ref. 67 (Copyright 2021 American Chemical Society).